\documentclass{ws-jcsc}

\usepackage{amssymb}
\usepackage{graphicx}

\newcommand{\be}{\begin{equation}}
\newcommand{\ee}{\end{equation}}
\newcommand{\bra}{\langle}
\newcommand{\ket}{\rangle}
\newcommand{\bea}{\begin{eqnarray}}
\newcommand{\eea}{\end{eqnarray}}
\newcommand{\dis}{\displaystyle}

\begin{document}

\markboth{Authors' Names}{Instructions for Typesetting 
Manuscripts (Condensed Title for the Paper)}

%
\catchline{}{}{}{}{}
%

\title{BAYESIAN INFERENCE OF STOCHASTIC VOLATILITY MODEL BY HYBRID MONTE CARLO}

\author{Tetsuya Takaishi$^\dag$}

\address{
         Hiroshima University of Economics,\\
         Hiroshima 731-0192  JAPAN \\
$^\dag$takaishi@hiroshima-u.ac.jp}

\maketitle

\begin{history}
\received{(Day Month Year)}
\revised{(Day Month Year)}
\accepted{(Day Month Year)}
\end{history}

\begin{abstract}
The hybrid  Monte Carlo (HMC) algorithm is 
applied for the Bayesian inference of the stochastic volatility (SV) model.
We use the HMC algorithm for the Markov chain Monte Carlo updates 
of volatility variables of the SV model.
First we compute parameters of the SV model by using  the artificial financial data and 
compare the results from the HMC algorithm with those from the Metropolis algorithm.
We find that the HMC algorithm decorrelates the volatility variables
faster than the Metropolis algorithm.
Second we make an empirical study for the time series of the Nikkei 225 stock index by the HMC algorithm.
We find the similar correlation behavior for the sampled data to the results from 
the artificial financial data and obtain a $\phi$ value close to one ($\phi \approx 0.977$), which means
that the time series has the strong persistency of the volatility shock.  

\end{abstract}


\section{Introduction}
Many empirical studies of financial prices such as stock indexes, exchange rates have confirmed 
that financial time series of price returns 
shows various interesting properties which can not be
derived from a simple assumption that the price returns follow the geometric Brownian motion.
Those properties are now classified as stylized facts\cite{Stanley,CONT}.
Some examples of the stylized facts are 
(i) fat-tailed distribution of return
(ii) volatility clustering 
(iii) slow decay of the autocorrelation time of the absolute returns.
The true dynamics behind the stylized facts is not fully understood. 
In order to imitate the real financial markets and to understand the origins of the stylized facts, 
a variety of models have been proposed and examined. 
Actually many models are able to capture some of the stylized facts\cite{Stauffer}-\cite{SPIN6}. 

In empirical finance the volatility is an important value to measure the risk.
One of the stylized facts of the volatility is that  
the volatility of price returns changes in time and shows clustering, so called "volatility clustering".
Then the histogram of the resulting price returns shows a fat-tailed distribution which 
indicates that the probability of having a large price change is higher  
than that of the Gaussian distribution.
In order to mimic these empirical properties of the volatility 
and to forecast the future volatility values,
Engle advocated the autoregressive conditional hetroskedasticity (ARCH) model\cite{ARCH} 
where the volatility variable changes deterministically depending on the past squared value of the return. 
Later the ARCH model is generalized by 
adding also the past volatility dependence to the volatility change. 
This model is known as the generalized ARCH (GARCH) model\cite{GARCH}.
The parameters of the GARCH model applied to financial time series 
are conventionally determined by the maximum likelihood method.
There are many extended versions of GARCH models, 
such as EGARCH\cite{EGARCH}, GJR\cite{GJR}, QGARCH\cite{QGARCH1,QGARCH2} models etc., 
which are designed to increase the ability to forecast the volatility value. 

The stochastic volatility (SV) model\cite{SVMCMC1,SV} is another model 
which captures the properties of the volatility.  
In contrast to the GARCH model,
the volatility of the SV model changes stochastically in time.
As a result the likelihood function of the SV model is given as 
a multiple integral of the volatility variables.
Such an integral in general is not analytically calculable 
and thus the determination of the parameters of the SV model 
by the maximum likelihood method becomes difficult.
To overcome this difficulty in the maximum likelihood method  
the Markov Chain Monte Carlo (MCMC) method based on the Bayesian approach is proposed and developed\cite{SVMCMC1}.
In the MCMC of the SV model one has to update not only the parameter variables but also the volatility ones 
from a joint probability distribution of the parameters and the volatility variables. 
The number of the volatility variables to be updated increases  with the data size of time series.
The first proposed update scheme of the volatility variables is based on the local update such as the Metropolis-type 
algorithm\cite{SVMCMC1}. 
It is however known that 
when the local update scheme is used for the volatility variables 
having interactions to their neighbor variables in time, 
the autocorrelation time of sampled volatility variables becomes large and 
thus the local update scheme becomes ineffective\cite{SVMCMC2}.
In order to improve the efficiency of the local update method 
the blocked scheme which updates several variables at once is also proposed\cite{SVMCMC2,Watanabe}.
A recent survey on the MCMC studies of the SV model is seen in Ref.25.

In our study we use the HMC algorithm\cite{HMC} 
which had not been considered seriously for the MCMC simulation of the SV model.
In finance there exists an application of the HMC algorithm to the GARCH model\cite{HMCGARCH}
where three GARCH parameters are updated by the HMC scheme.
It is more interesting to apply the HMC for updates of the volatility variables
because  the HMC algorithm is a global update scheme which can update
all variables at once. This feature of the HMC algorithm can be used for 
the global update of the volatility variables which can not be achieved by 
the standard Metropolis algorithm.  
A preliminary study\cite{ICIC2008} shows that the HMC algorithm samples the volatility variables 
effectively.
In this paper we give a detailed description of the HMC algorithm and 
examine the HMC algorithm with artificial financial data up to the data size of T=5000. 
We also make an empirical analysis of the Nikkei 225 stock index by 
the HMC algorithm.

\section {Stochastic Volatility Model}

The standard version of the SV model\cite{SVMCMC1,SV} is given by 
\be
y_t = \sigma_t \epsilon_t = \exp(h_t/2)\epsilon_t,
\label{eq:SV}
\ee
\be
h_t = \mu +\phi (h_{t-1} -\mu) +\eta_t,
\ee
where $y_t=(y_1,y_2,...,y_n)$ represents the time series data, $h_t$ is defined 
by $h_t=\ln \sigma_t^2$ 
and $\sigma_t$ is called volatility.
We also call $h_t$ volatility variable.
The error terms $\epsilon_t$ and $\eta_t$ are taken from independent normal distributions
$N(0,1)$ and $N(0,\sigma_\eta^2)$ respectively.
We assume that $|\phi|<1$.
When $\phi$ is close to one, the model exhibits the strong persistency of the volatility shock.

For this model the parameters to be determined are $\mu$, $\phi$ and $\sigma^2_\eta$.
Let us use $\theta$ as  $\theta=(\mu, \phi,\sigma^2_\eta)$.
Then the likelihood function $L({\bf \theta})$ for the SV model is written as 
\be
L({\bf \theta})=\int \prod_{t=1}^n f(\epsilon_t|\sigma_t^2) f(h_t|\theta) dh_1dh_2...dh_n,
\label{LFUNC}
\ee
where
\be
f(\epsilon_t|\sigma_t^2)=\left(2\pi \sigma_t^2\right)^{-\frac12}\exp\left(-\frac{y_t^2}{2\sigma_t^2}\right),
\ee
\be
f(h_1|\theta)=\left(\frac{2\pi \sigma_\eta^2}{1-\phi^2}\right)^{-\frac12}  
\exp\left(-\frac{[h_1-\mu]^2}{2\sigma_\eta^2/(1-\phi^2)}\right),
\ee

\be
f(h_t|\theta)=\left(2\pi\sigma_\eta^2\right)^{-\frac12} 
\exp\left(-\frac{[h_t-\mu-\phi(h_{t-1}-\mu)]^2 }{2\sigma_\eta^2}\right).
\ee
As seen in Eq.(\ref{LFUNC}), $L({\bf \theta})$ is constructed as a multiple integral of the volatility variables.
For such an integral it is difficult to apply the maximum likelihood method 
which estimates values of $\theta$ by maximizing the likelihood function. 
Instead of using the maximum likelihood method  
we perform the MCMC simulations based on the  Bayesian inference as explained in the next section. 

\section{Bayesian inference for the SV model}
From the Bayes' rule, the probability distribution of the parameters $\theta$
is given by
\be
f(\theta|y)=\frac1Z L({\bf \theta}) \pi(\bf \theta),
\ee
where $Z$ is the normalization constant $Z=\int L({\bf \theta}) \pi({\bf \theta}) d\theta$ and
$\pi(\bf \theta)$ is a prior distibution of ${\bf \theta}$ for which we make a certian assumption.
The values of the parameters are inferred as the expectation values of  
$\theta$ given by 
\be
\bra {\bf \theta} \ket = \int {\bf \theta} f(\theta|y) d\theta.
\ee
In general this integral can not be performed analytically.
For that case, one can use the MCMC method to estimate the expectation values numerically. 

In the MCMC method, we first generate a series of $\theta$  with 
a probability of $P(\theta)= f(\theta|y)$.
Let $\theta^{(i)}=(\theta^{(1)},\theta^{(2)},...,\theta^{(k)})$ be values of $\theta$ generated  
by the MCMC sampling. Then using these $k$ values the expectation value of $\theta$ is estimated by
an average as
\be
\bra {\bf \theta} \ket = \frac1k\sum_{i=1}^k \theta^{(i)}.
\ee
The statistical error for $k$ independent samples is proportional to $\dis \frac1{\sqrt{k}}$.
When the sampled data are correlated the statistical error will be proportional to 
$\dis \sqrt{\frac{2\tau}{k}}$ where $\tau$ is the autocorrelation time between the sampled data.
The value of $\tau$ depends on the MCMC sampling scheme we take.
In order to reduce the statistical error within limited sampled data
it is better to choose an MCMC method which is able to generate data with a small $\tau$.

\subsection{MCMC Sampling of $\theta$}
For the SV model, in addition to $\theta$, volatility variables $h_t$ also
have to be updated since they should be integrated out as in Eq.(\ref{LFUNC}). 
Let $P(\theta,h_t)$ be the joint probability distribution of
$\theta$ and $h_t$.
Then $P(\theta,h_t)$ is given by
\be 
P(\theta,h_t) \sim \bar{L}(\theta,h_t)\pi(\theta),
\label{eq:prob}
\ee
where 
\be
\bar{L}(\theta,h_t) = \prod_{t=1}^n f(\epsilon_t|h_t) f(h_t|\theta).
\ee
For the prior $\pi(\theta)$ we assume that $\pi(\sigma_\eta^2)\sim (\sigma_\eta^2)^{-1}$
and for others $\pi(\mu) =\pi(\phi) =constant.$

The MCMC sampling methods for $\theta$ are given in the following\cite{SVMCMC1,SV}.  
The probability distribution for each parameter 
can be derived from Eq.(\ref{eq:prob}) by extracting the part including the corresponding parameter. 

\begin{itemize}
\item $\sigma_\eta^2$ update scheme.

The probability distribution of $\sigma_\eta^2$ is given by
\be
P(\sigma_\eta^2)\sim (\sigma_\eta^2)^{-\frac{n}2-1} \exp\left(-\frac{A}{\sigma_\eta^2}\right),
\label{sigma}
\ee
where 
\be
A=\frac12\{(1-\phi^2)(h_1-\mu)^2+\sum_{t=2}^n [h_t-\mu -\phi(h_{t-1}-\mu)]^2\}.
\ee
Since Eq.(\ref{sigma}) is an inverse gamma distribution 
we can easily draw a value of $\sigma_\eta^2$  
by using an appropriate statistical library in the computer.

\item $\mu$ update scheme.

The probability distribution of $\mu$ is given by
\be
P(\mu) \sim \exp\left\{-\frac{B}{2\sigma_\eta^2}(\mu-\frac{C}{B})^2 \right\},
\label{mu}
\ee
where 
\be
B=(1-\phi^2)+(n-1)(1-\phi)^2, 
\ee
and 
\be
C=(1-\phi^2)h_1+(1-\phi)\sum_{t=2}^n (h_t-\phi h_{t-1}).
\ee
$\mu$ is drawn from a Gaussian distribution of Eq.(\ref{mu}).

\item $\phi$ update scheme.

The probability distribution of $\phi$ is given by
\be
P(\phi)\sim (1-\phi^2)^{1/2}\exp\{-\frac{D}{2\sigma_\eta^2}(\phi-\frac{E}{D})^2\},
\label{phi}
\ee
where 
\be
D=-(h_1-\mu)^2+\sum_{t=2}^n(h_{t-1}-\mu)^2$, and $E=\sum_{t=1}^n (h_t-\mu)(h_{t-1}-\mu).
\ee
In order to update $\phi$ with Eq.(\ref{phi}), we use the Metropolis-Hastings algorithm\cite{METRO,MH}.
Let us write Eq.(\ref{phi}) as
$ \dis P(\phi)\sim P_1(\phi)P_2(\phi)$ where
\be
P_1(\phi)= (1-\phi^2)^{1/2},
\ee
\be
P_2(\phi) \sim \exp\{-\frac{D}{2\sigma_\eta^2}(\phi-\frac{E}{D})^2\}.
\label{phi2}
\ee
Since $P_2(\phi)$ is a Gaussian distribution we can easily draw $\phi$ from Eq.(\ref{phi2}).
Let $\phi_{new}$ be a candidate given from Eq.(\ref{phi2}).
Then in order to obtain the correct distribution,
$\phi_{new}$ is accepted with the following probability $P_{MH}$.
\be
P_{MH}=\min\left\{\frac{P(\phi_{new})P_2(\phi)}{P(\phi)P_2(\phi_{new})},1 \right\}
=\min\left\{\sqrt{\frac{(1-\phi^2_{new})}{(1-\phi^2)}},1\right\}.
\ee
In addition to the above step we restrict $\phi$ within $[-1,1]$ to avoid a negative value 
in the calculation of square root.

\end{itemize}

\subsection{Probability distribution for $h_t$}

The probability distribution of the volatility variables $h_t$ 
is given by   
\bea 
\label{eq:ham}
& P(h_t)\equiv P(h_1,h_2,...,h_n)  \sim \vspace{2cm}  \\ \nonumber
& \exp \left(-\sum_{i=1}^n \{\frac{h_t}{2}+\frac{\epsilon_t^2}{2}e^{-h_t}\}
-\frac{[h_1-\mu]^2}{2\sigma_\eta^2/(1-\phi^2)}
-\sum_{i=2}^n \frac{[h_t-\mu-\phi(h_{t-1}-\mu)]^2}{2\sigma_\eta^2}\right).
\eea
This probability distribution is not a simple function 
to draw values of $h_t$.
A conventional method is the Metropolis method\cite{METRO,MH} which updates the variables
locally. There are several methods\cite{SVMCMC1,SV,SVMCMC2,Watanabe} developed to update $h_t$ from Eq.(\ref{eq:ham}).
Here we use the HMC algorithm to update $h_t$ globally.
The HMC algorithm is described in the next section.

\section{Hybrid Monte Carlo Algorithm}
Originally the HMC algorithm is developed for 
the MCMC simulations of the lattice Quantum Chromo Dynamics (QCD) calculations\cite{HMC}.
A major difficulty of the lattice QCD calculations is the inclusion of
dynamical fermions. The effect of the dynamical fermions is incorporated by
the determinant of the fermion matrix. 
The computational work of the determinant calculation requires
$O(V^3)$ arithmetic operations\cite{UKAWA}, where $V$ is the volume of a 4-dimensional lattice.
A typical size of the volume is $V > 10^4$. 
The standard Metropolis algorithm which locally updates variables 
does not work since each local update requires $O(V^3)$ arithmetic operations for a determinant calculation,
which results in unacceptable computational cost in total. 
Since the HMC algorithm is a global update method, the computational cost remains in 
the acceptable region.

The basic idea of  the HMC algorithm is a combination of molecular dynamics (MD) simulation 
and Metropolis accept/reject step.
Let us consider to evaluate the following expectation value $\bra O(x) \ket$ by the HMC algorithm.
\be
\bra O(x) \ket   =  \int O(x) f(x) dx   =  \int O(x) e^{ln f(x)} dx,
\label{eq:integral}
\ee
where $x=(x_1,x_2,...,x_n)$, $f(x)$ is a probability density
and 
$O(x)$ stands for an function of $x$.
First we introduce momentum variables $p=(p_1,p_2,...,p_n)$ conjugate to the variables $x$  and then 
rewrite Eq.(\ref{eq:integral}) as
\be 
\bra O(x) \ket   =   \frac1{Z}  \int O(x) e^{- \frac12 p^2 + ln f(x)} dxdp  =   \frac1{Z}  \int O(x) e^{- H(p,x)} dxdp.
\ee
where $Z$ is a normalization constant given by
\be
Z= \int \exp\left(-\frac12 p^2\right)dp,
\ee
and  $p^2$ stands for $\sum_{i=1}^n p_i^2$.
$H(p,x)$ is the Hamiltonian defined by 
\be
H(p,x) = \frac12 p^2 - ln f(x).
\ee
Note that the introduction of $p$ does not change the value of $\bra O(x) \ket$.

In the HMC algorithm, new candidates of the variables $(p,x)$ are drawn by 
integrating the Hamilton's equations of motion,
\bea
& \frac{\dis dx_i}{\dis dt}  = & \frac{\partial H}{\partial p_i},\\
& \frac{\dis dp_i}{\dis dt}  = & -\frac{\partial H}{\partial x_i}.
\eea
In general the Hamilton's equations of motion are not solved analytically.
Therefore we solve them  numerically by doing the MD simulation.
Let $T_{MD}(\Delta t)$ be an elementary MD step 
with a step size $\Delta t$, 
which evolves $(p(t),x(t))$ to $(p(t+\Delta t),x(t+\Delta t))$:

\be
T_{MD}(\Delta t):(p(t),x(t)) \rightarrow (p(t+\Delta t),x(t+\Delta t)).
\ee

Any integrator can be used for the MD simulation provided that
the following conditions are satisfied\cite{HMC}

\begin{itemize}
\item area preserving
\be
dp(t)dx(t)dx=dp(t+\Delta t)dx(t+\Delta t).
\ee
\item time reversibility
\be
T_{MD}(-\Delta t):(p(t+\Delta t),x(t+\Delta t)) \rightarrow (p(t),x(t)).
\ee
\end{itemize}

The simplest and often used integrator satisfying the above two conditions  
is the 2nd order leapfrog integrator given by
\bea
& x_i(t+\Delta t/2)& = x_i(t)+\frac{\Delta t}{2}p_i(t) \nonumber \\
& p_i(t+\Delta t)& = p(t)_i-\Delta t\frac{\partial H}{\partial x_i} \nonumber \\
& x_i(t+\Delta t)& = x_i(t+\Delta t/2)+\frac{\Delta t}{2}p_i(t+\Delta t).
\label{leapfrog}
\eea
In this study we use this integrator.
The numerical integration is performed $N$ steps repeatedly by Eq.(\ref{leapfrog}) 
and in this case the total trajectory length $\lambda$ of the MD is  $\lambda = N \times \Delta t$. 

At the end of the trajectory we obtain new candidates  $(p^\prime,x^\prime)$.
These candidates are accepted with the Metropolis test, i.e. $(p^\prime,x^\prime)$ are globally accepted 
with the following probability,
\be 
P = \min\{1,\frac{\exp\left(-H(p^\prime,x^\prime)\right)}{\exp\left(-H(p,x)\right)}\}=\min\{1,\exp\left(-\Delta H\right)\},
\ee
where $\Delta H$ is the energy difference given by $\Delta H = H(p^\prime,x^\prime) -H(p,x)$.
Since we integrate the Hamilton's equations of motion approximately by an integrator,
the total Hamiltonian is not conserved, i.e. $\Delta H \neq 0$.
The acceptance or the magnitude of $\Delta H$ is tuned by the step size $\Delta t$ 
to obtain a reasonable acceptance.
Actually there exists the optimal acceptance which 
is about $60-70\%$ for 2nd order integrators\cite{HOHMC,HOHMC2}.
Surprisingly the optimal acceptance is not dependent of the model we consider. 
For the n-th order integrator the optimal acceptance is expected to be\cite{HOHMC} 
$\dis \sim \exp\left(-\frac1n\right)$. 

We could also use higher order integrators 
which give us a smaller energy difference $\Delta H$. 
However the higher order integrators are not always effective 
since they need more arithmetic operations than the lower order integrators\cite{HOHMC,HOHMC2}.
The efficiency of the higher order integrators depends on the model we consider.
There also exist improved integrators which have less arithmetic operations 
than the conventional integrators\cite{MNHOMC}.

For the volatility variables $h_t$, from Eq.(\ref{eq:ham}),
the Hamiltonian can be defined by
\be
H(p_t,h_t)=\sum_{i=1}^n \frac12 p_i^2 +
\sum_{i=1}^n \{\frac{h_i}{2}+\frac{\epsilon_i^2}{2}e^{-h_i}\}
+\frac{[h_1-\mu]^2}{2\sigma_\eta^2/(1-\phi^2)}
+\sum_{i=2}^n \frac{[h_i-\mu-\phi(h_{i-1}-\mu)]^2}{2\sigma_\eta^2},
\ee
where $p_i$ is defined as a conjugate momentum to $h_i$.
Using this Hamiltonian we perform the HMC algorithm for updates of $h_t$.

\section{Numerical Studies}
In order to test the HMC algorithm 
we use artificial financial time series data generated 
by the SV model with a set of known parameters
and perform the MCMC simulations to the artificial financial data by the HMC algorithm. 
We also perform the MCMC simulations by the Metropolis algorithm 
to the same artificial data and
compare the results with those from the HMC algorithm.

\begin{figure}
\vspace{2mm}
\centering
\includegraphics[height=4.1cm]{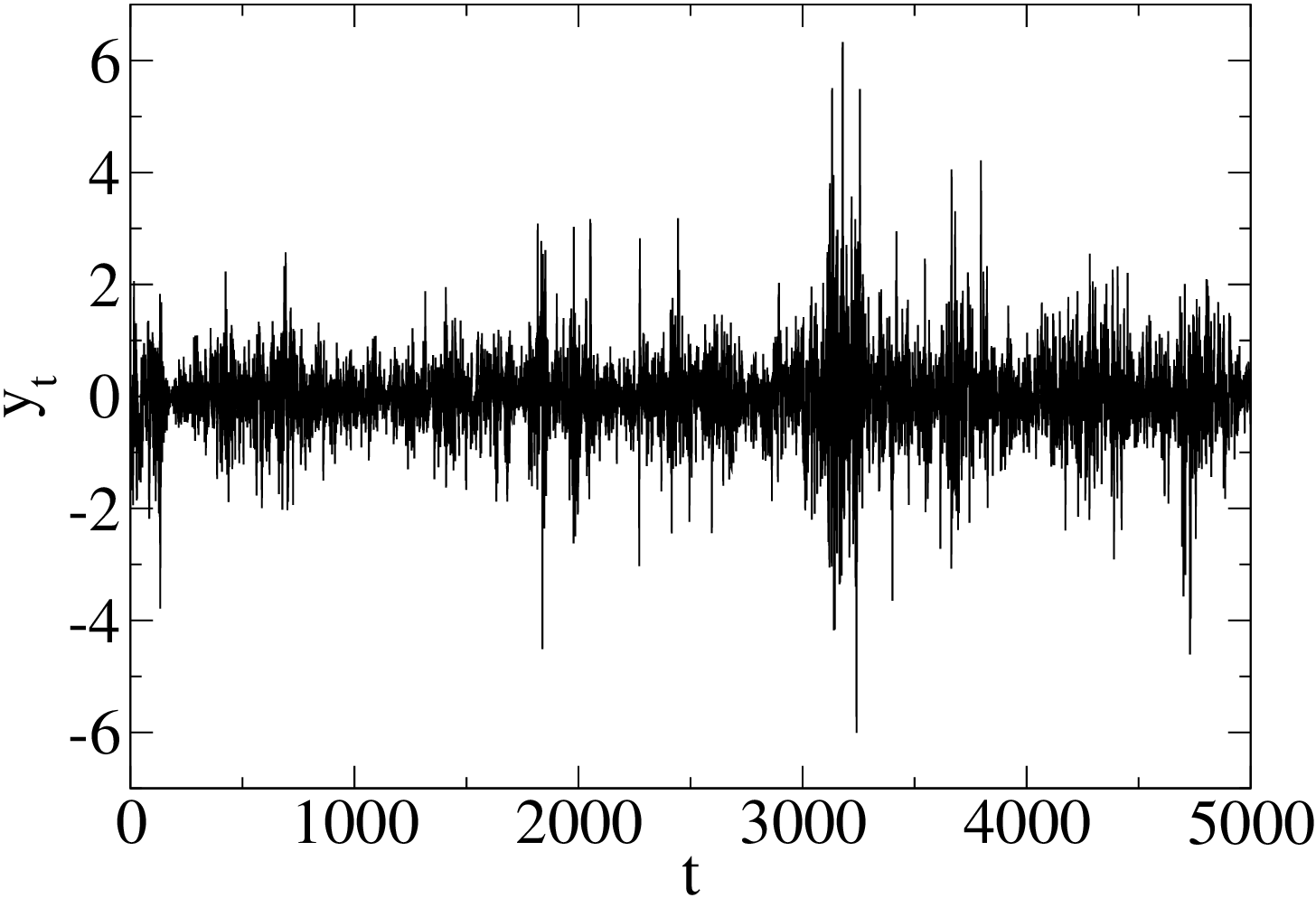}
\vspace{-1mm}
\caption{
The artificial SV time series used for this study.}
\label{fig:SVdata}
\vspace{-2mm}
\end{figure}

Using Eq.(\ref{eq:SV}) with $\phi=0.97$,$\sigma_{\eta}^2 = 0.05$ and $\mu=-1$ 
we have generated 5000 time series data. 
The time series generated by Eq.(\ref{eq:SV}) is shown in Fig.\ref{fig:SVdata}. 
From those data we prepared 3 data sets: (1)T=1000 data (the first 1000 of the time series),
(2)T=2000 data (the first 2000 of the time series)  and (3) T=5000 (the whole data). 
To these data sets we made the Bayesian inference by the HMC
and Metropolis algorithms. Precisely speaking both algorithms are used only for 
the MCMC update of the volatility variables. For the update of the SV parameters
we used the update schemes in Sec.3.1.

For the volatility update in the Metropolis algorithm,  
we draw a new candidate of the volatility variables randomly, i.e.
a new volatility $h^{new}_t$ is given from the previous value $h^{old}_t$ by
\be
h^{new}_t = h^{old}_t+\delta(r-0.5),
\ee   
where $r$ is a uniform random number in $[0,1)$
and $\delta$ is a parameter to tune the acceptance.
The new volatility $h^{new}_t$ is accepted with the acceptance $P_{metro}$ 
\be
P_{metro}=\min\left\{1,\frac{P(h^{new}_t)}{P(h^{old}_t)}\right\},
\ee
where $P(h_t)$ is given by Eq.(\ref{eq:ham}).

The initial parameters for the MCMC simulations are set to $\phi=0.5$,$\sigma_{\eta}^2 = 1.0$ and $\mu=0$.
The first 10000 samples are discarded as thermalization or burn-in process.
Then 200000 samples are recorded for analysis. 
The total trajectory length $\lambda$ of the HMC algorithm is set to $\lambda=1$
and the step size $\Delta t$ is tuned so that  
the acceptance of the volatility variables becomes more than 50\%.

\begin{figure}
\vspace{2mm}
\centering
\includegraphics[height=4.1cm]{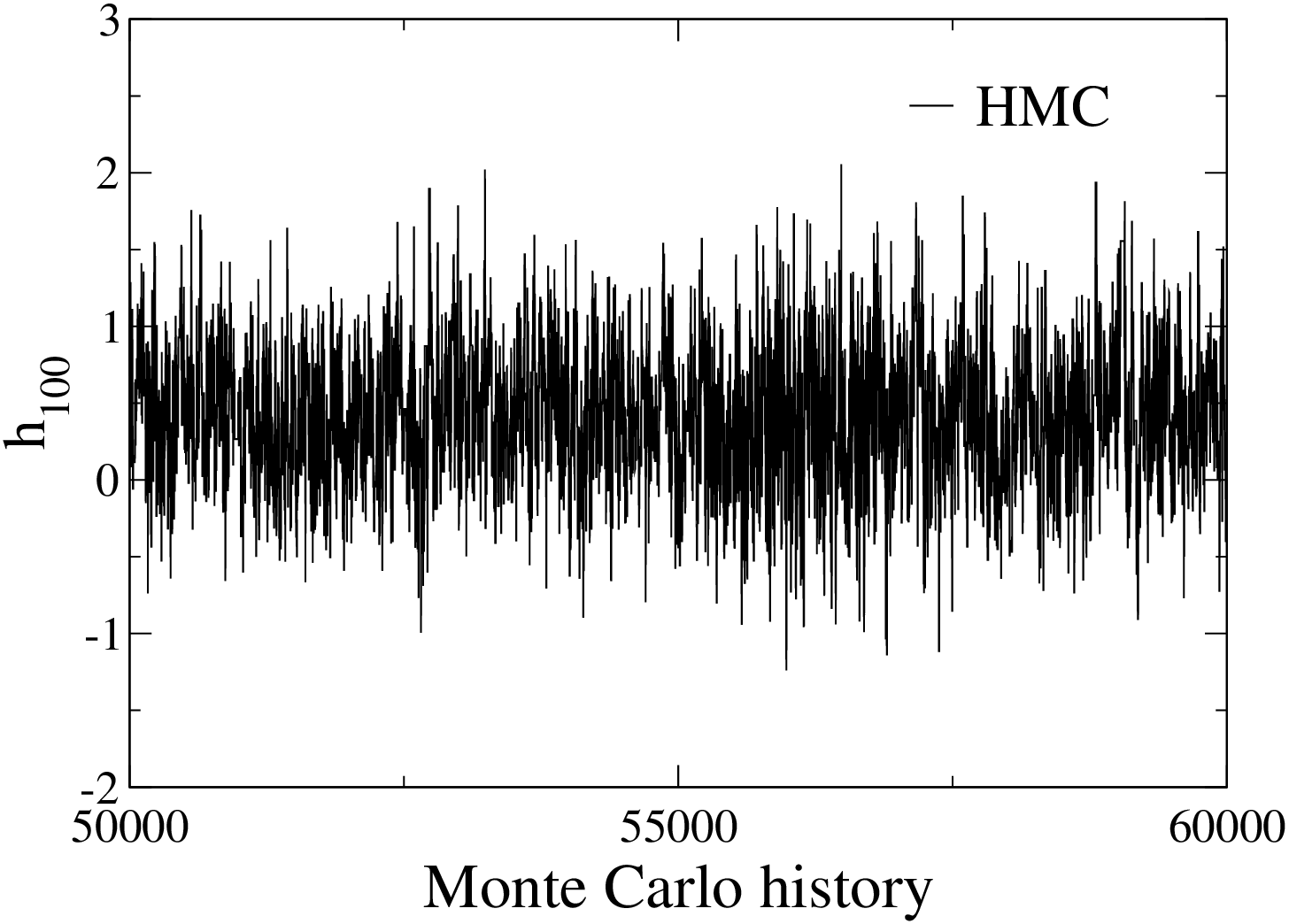}
\hspace{2mm}
\includegraphics[height=4.1cm]{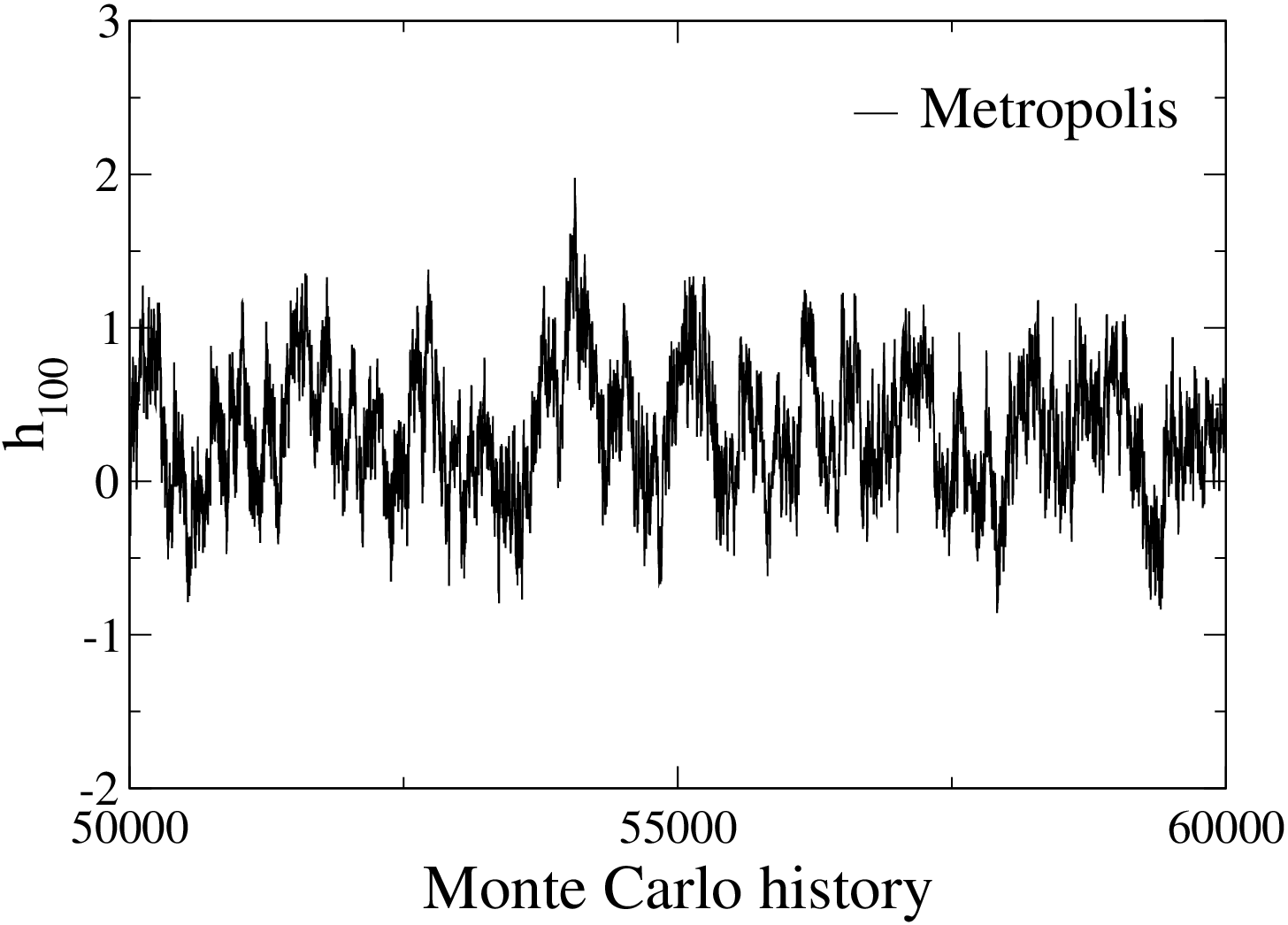}
\vspace{-2mm}
\caption{
Monte Carlo histories of $h_{100}$ generated by HMC (left) and Metropolis (right) 
with $T=2000$ data set.
The Monte Carlo histories in the window from 50000 to 60000 are shown.}
\label{fig:traj}
\vspace{-2mm}
\end{figure}

\begin{figure}
\vspace{1mm}
\centering
\includegraphics[height=4.2cm]{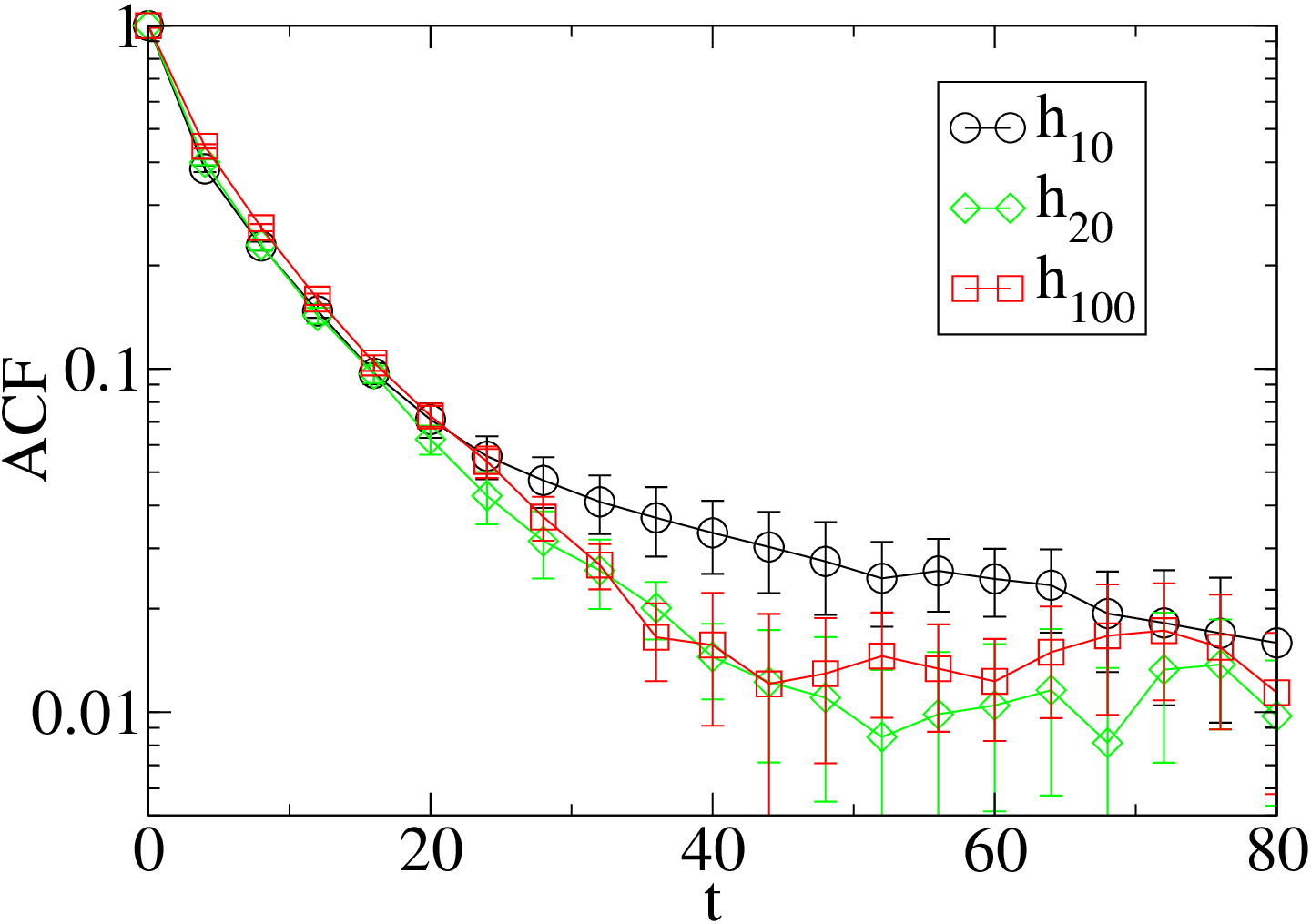}
\vspace{-2mm}
\caption{
Autocorrelation functions of three volatility variables $h_{10},h_{20}$ and $h_{100}$
sampled by the HMC algorithm for $T=2000$ data set. 
These autocorrelation functions show the similar behavior.}
\label{fig:threeauto}
\vspace{-1mm}
\end{figure}

First we analyze the sampled volatility variables.
Fig.\ref{fig:traj} shows the Monte Carlo (MC) history of the volatility variable $h_{100}$ from $T=2000$ data set. 
We take $h_{100}$ as the representative one of the volatility variables 
since we have observed the similar behavior for other volatility variables.
See also Fig.\ref{fig:threeauto} for the similarity of the autocorrelation functions of the volatility variables.  

A comparison of the volatility histories in Fig.\ref{fig:traj}
clearly indicates that
the correlation of the volatility variable sampled from the HMC algorithm is smaller than 
that from the Metropolis algorithm. 
To quantify this we calculate the autocorrelation function (ACF) of the volatility variable.
The ACF is defined as 
\be
ACF(t) = \frac{\frac1N\sum_{j=1}^N(x(j)- \bra x\ket )(x(j+t)-\bra x\ket)}{\sigma^2_x},
\ee
where $\bra x\ket$ and $\sigma^2_x$ are the average value and the variance of $x$ respectively.

Fig.\ref{fig:threeauto} shows the ACF for three volatility variables, $h_{10},h_{20}$ and $h_{100}$
sampled by the HMC.
It is seen that those volatility variables have the similar correlation behavior.
Other volatility variables also show the similar behavior.
Thus hereafter we only focus on the volatility variable $h_{100}$ as the representative one. 

\begin{figure}
\vspace{1mm}
\centering
\includegraphics[height=4.2cm]{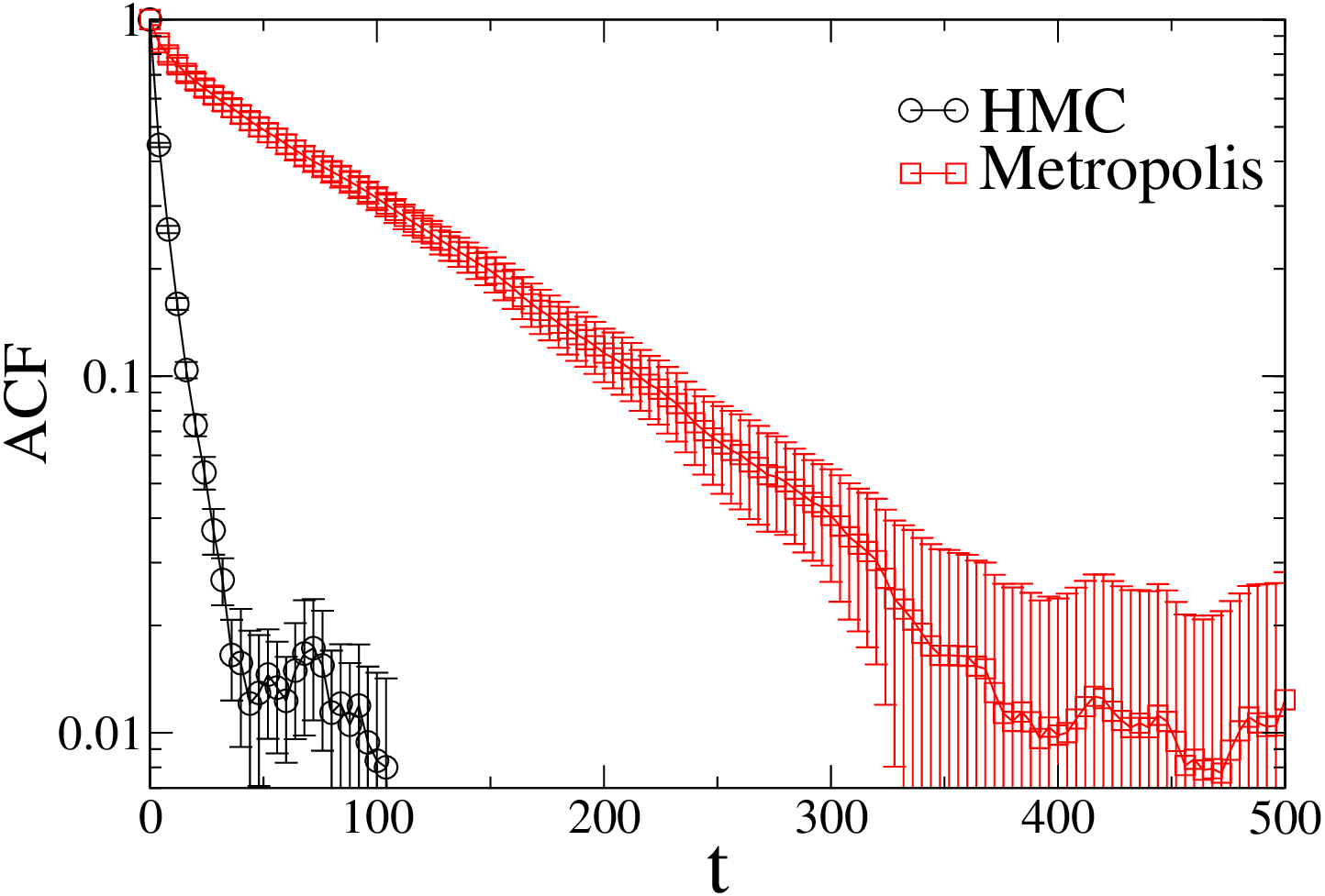}
\vspace{-2mm}
\caption{
Autocorrelation function of the volatility variable $h_{100}$
by the HMC and Metropolis algorithms for $T=2000$ data set.}
\label{fig:auto}
\vspace{-1mm}
\end{figure}

Fig.\ref{fig:auto} compares the ACF of $h_{100}$ by the HMC and Metropolis algorithms. 
It is obvious that the ACF by the HMC decreases more rapidly than that by the Metropolis algorithm.
We also calculate the autocorrelation time $\tau_{int}$ defined by
\be
\tau_{int} = \frac12 + \sum_{t=1}^{\infty} ACF(t).
\ee
The results of $\tau_{int}$ of the volatility variables are given in Table 1.
The values in the parentheses represent the statistical errors estimated by the jackknife method.
We find that the HMC algorithm gives a smaller autocorrelation time than the Metropolis algorithm,
which means that the HMC algorithm samples the volatility variables more effectively 
than the Metropolis algorithm.

\begin{table}
\centering
\begin{tabular}{cc|cccc}
&             & \makebox[20mm]{$\phi$}   & \makebox[20mm]{$\mu$}  &  \makebox[20mm]{$\sigma^2_{\eta}$}   &  \makebox[20mm]{$h_{100}$}    \\  \hline
&true         & 0.97     &  -1    &  0.05      &     \\ \hline \hline
T=1000&HMC    &  0.973   & -1.13  &  0.053     &     \\
& SD          &  0.010   &  0.51  &  0.017     &     \\ 
& SE          &  0.0004  &  0.003 &  0.001     &     \\
&$2\tau_{int}$ &  360(80) & 3.1(5) &  820(200)  &   12(1)   \\ \hline 
&Metropolis   &  0.973   & -1.14  &  0.053     &           \\
& SD          &  0.011   &  0.40  &  0.017     &     \\
& SE          &  0.0005  &  0.003 &  0.0013    &     \\
&$2\tau_{int}$ & 320(60)  & 10.1(8)&  720(160)  &   190(20) \\ \hline \hline

T=2000&HMC    &  0.978   & -0.92  &  0.053     &     \\
& SD          &  0.007   &  0.26  &  0.012     &     \\ 
& SE          &  0.0003  &  0.001 &  0.0009    &     \\
&$2\tau_{int}$ &  540(60) & 3(1)   &  1200(150) &   18(1)   \\ \hline
&Metropolis   &  0.978   & -0.92  &  0.052     &     \\
& SD          &  0.007   &  0.26  &  0.011     &     \\
& SE          &  0.0003  &  0.003 &  0.0009    &     \\
&$2\tau_{int}$ & 400(100) & 13(2)  &  1000(270) &   210(50) \\ \hline \hline

T=5000&HMC    &  0.969   & -1.00  &  0.056     &     \\
& SD          &  0.005   &  0.11  &  0.009     &     \\
& SE          &  0.0003  &  0.0004&  0.0007    &     \\
&$2\tau_{int}$ &  670(100) & 4.2(7)   &  1250(170) &   10(1)   \\ \hline
&Metropolis   &  0.970   & -1.00  &  0.054     &     \\
& SD          &  0.005   &  0.12  &  0.008     &     \\
& SE          &  0.00023 &  0.0011&  0.0005    &     \\
&$2\tau_{int}$ & 510(90) & 30(10)  &  960(180) &   230(28)

\end{tabular}
\caption{Results estimated by the HMC and Metropolis algorithms. {\bf SD} stands for Standard Deviation and
{\bf SE} stands for Statistical Error. The statistical errors are estimated by the jackknife method.
We observe no significant differences on the autocorrelation times among three data sets.
}
\vspace{-2mm}
\end{table}

\begin{figure}
\centering
\includegraphics[height=4.1cm]{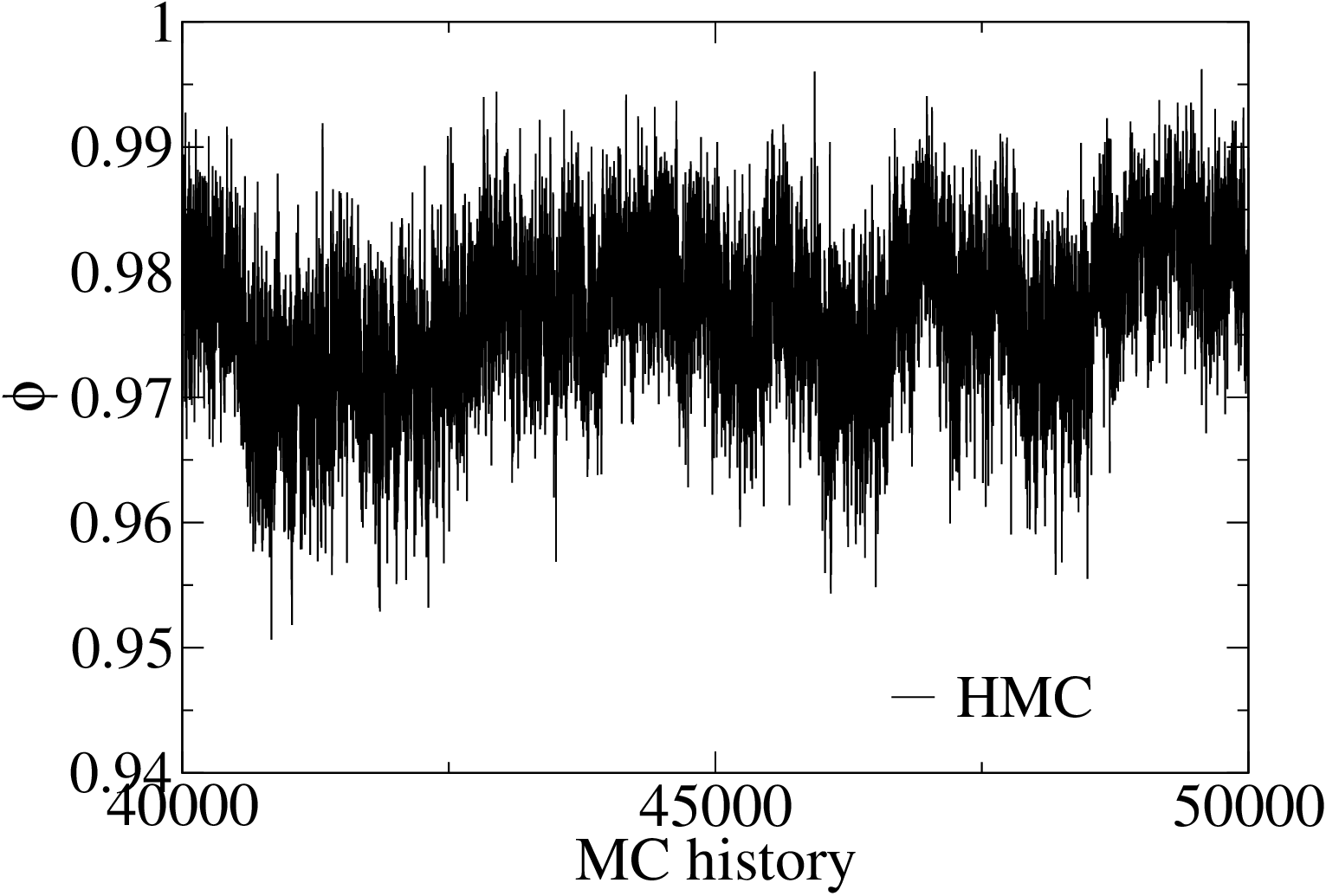}
\includegraphics[height=4.1cm]{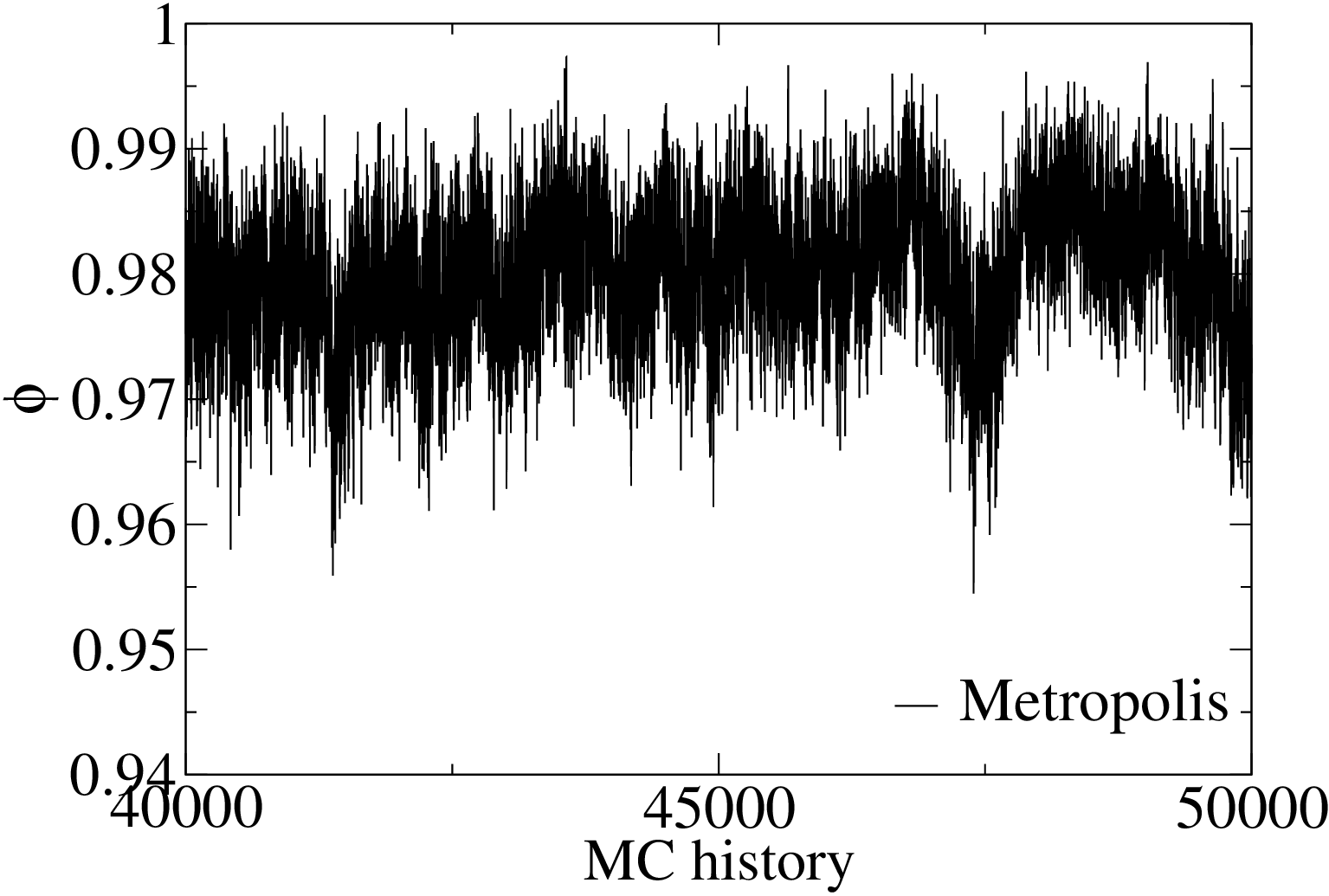}
\vspace{-2mm}
\caption{
Monte Carlo histories of $\phi$ generated by HMC (left) and Metropolis (right)
for $T=2000$ data set.}
\label{fig:histphi}
\vspace{-1mm}
\end{figure}

\begin{figure}
\centering
\includegraphics[height=4.1cm]{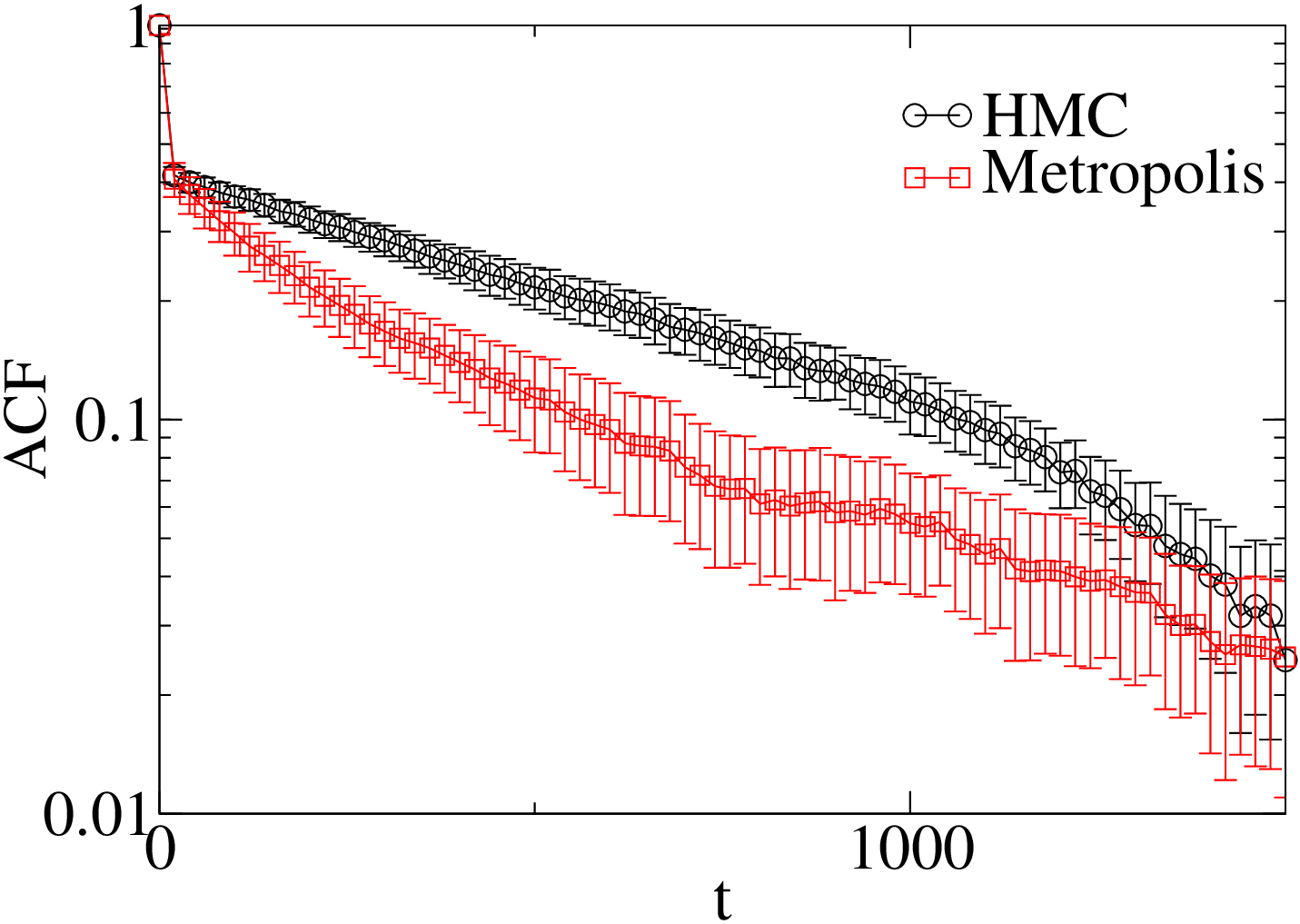}
\includegraphics[height=4.1cm]{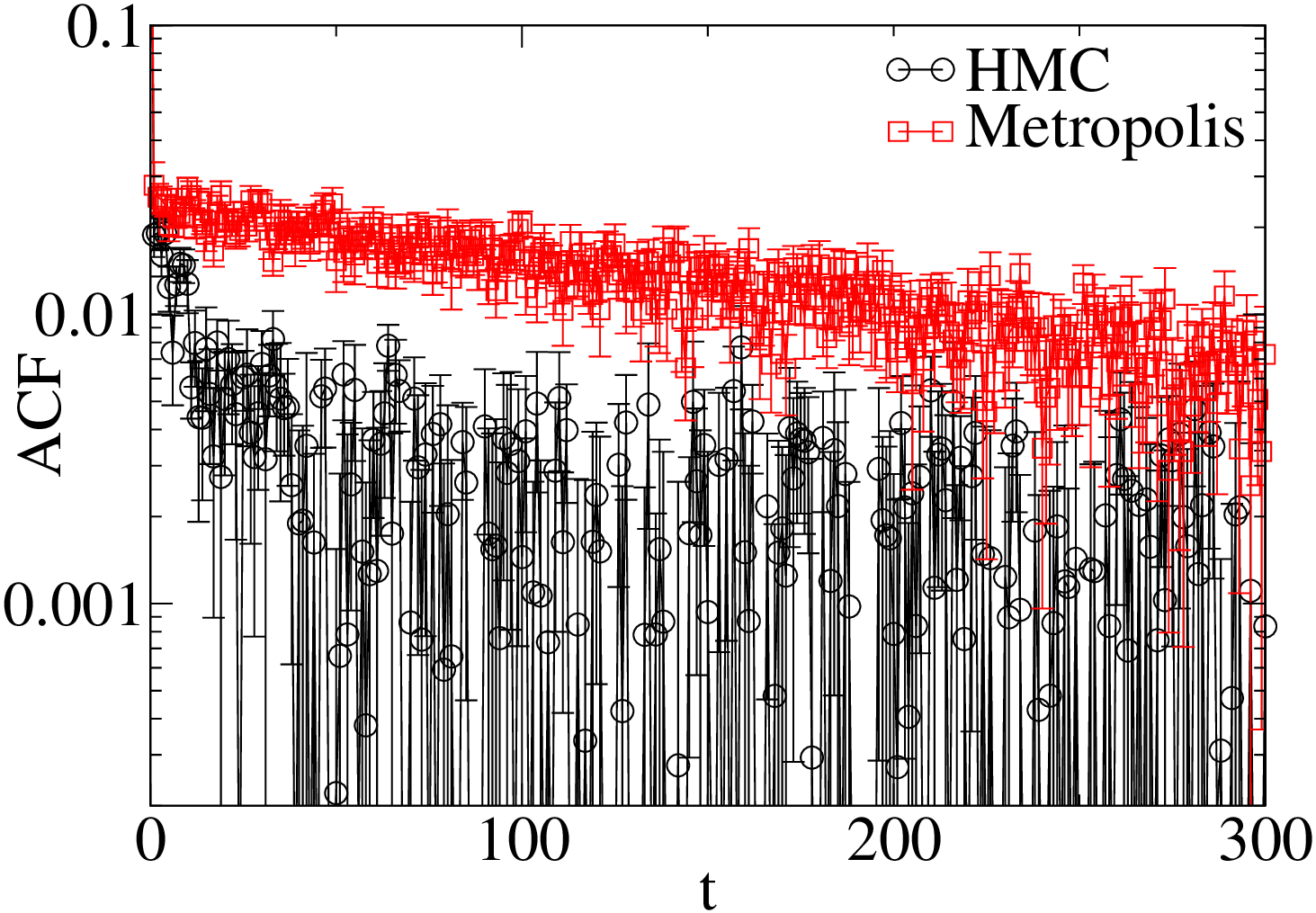}
\vspace{-2mm}
\caption{
Autocorrelation functions of $\phi$(left) and $\mu$(right) 
by the HMC and Metropolis algorithm for $T=2000$ data set.}
\label{fig:autophimu}
\vspace{-1mm}
\end{figure}

Next we analyze the sampled SV parameters.
Fig.\ref{fig:histphi} shows MC histories of the $\phi$ parameter sampled by the HMC and Metropolis algorithms.
It seems that both algorithms have the similar correlation for $\phi$.  
This similarity is also seen in the ACF in Fig.\ref{fig:autophimu}(left), i.e. 
both autocorrelation functions decrease in the similar rate with time $t$. 
The autocorrelation times of $\phi$ are very large as seen in Table 1.
We also find the similar behavior for $\sigma_\eta^2$, i.e. both autocorrelation times of $\sigma_\eta^2$
are large.  

On the other hand we see small autocorrelations for $\mu$ as seen in Fig.\ref{fig:autophimu}(right).
Furthermore we observe that the HMC algorithm gives a smaller $\tau_{int}$ for $\mu$ than that of the Metropolis algorithm,
which means that HMC algorithm samples $\mu$ more effectively than the Metropolis algorithm
although the values of $\tau_{int}$ for $\mu$ take already very small even for the Metropolis algorithm.

The values of the SV parameters estimated by the HMC and the Metropolis algorithms are listed in Table 1.
The results from both algorithms well reproduce the true values used for the generation of the artificial financial data.
Furthermore for each parameter and each data set, the estimated parameters by the HMC and the Metropolis algorithms agree well.
And their standard deviations also agree well. 
This is not surprising because the same artificial financial data, 
thus the same likelihood function is used for both MCMC simulations by the HMC and Metropolis algorithms.
Therefore they should agree each other.

\section{Empirical Analysis}
\begin{figure}
\vspace{1mm}
\centering
\includegraphics[height=4.1cm]{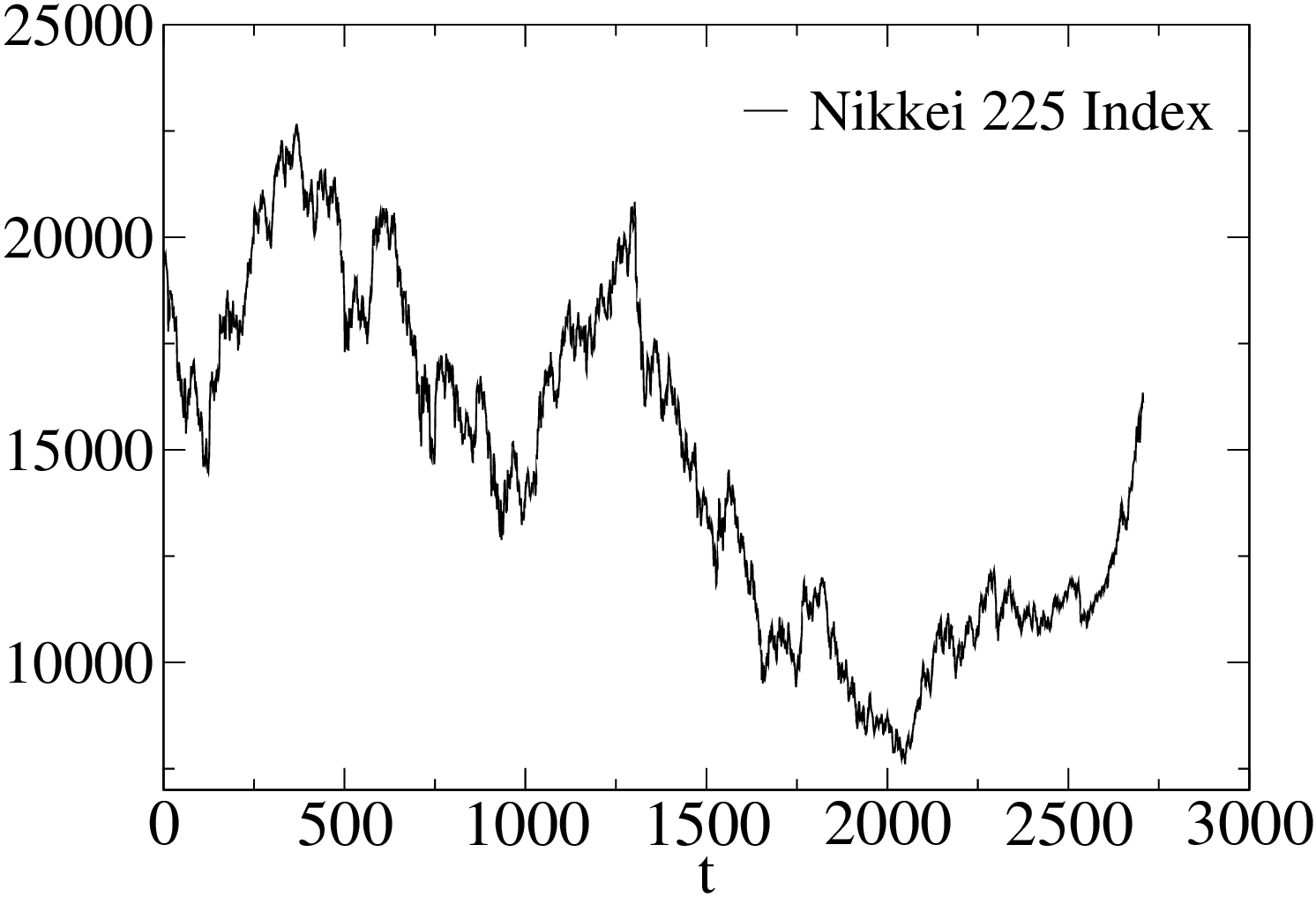}
\includegraphics[height=4.05cm]{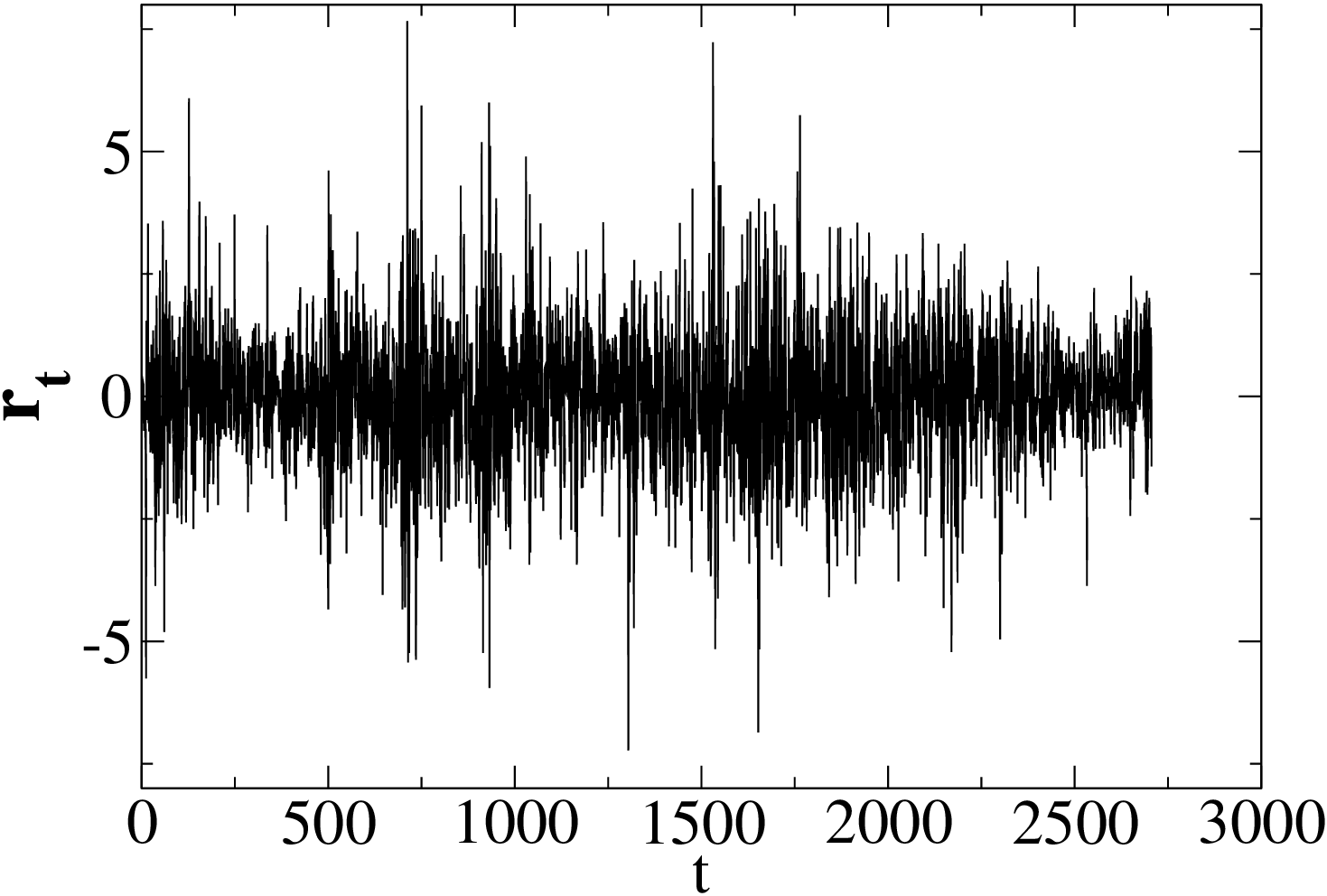}
\vspace{-2mm}
\caption{
Nikkei 225 stock index from 4 January 1995  to 30 December 2005(left) and returns(right).}
\label{fig:nikkei}
\vspace{-1mm}
\end{figure}

\begin{figure}
\centering
\includegraphics[height=4.1cm]{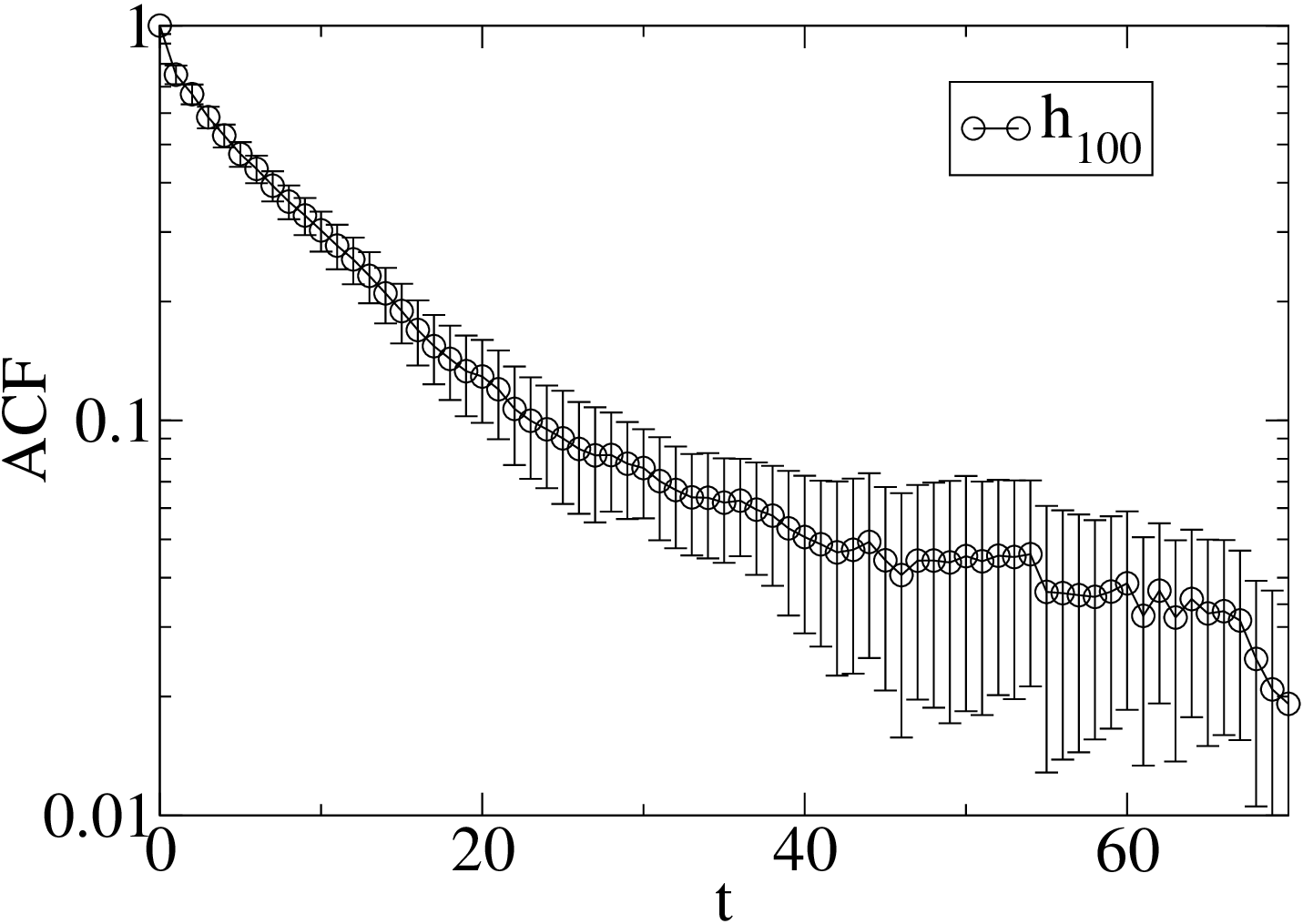}
\includegraphics[height=4.0cm]{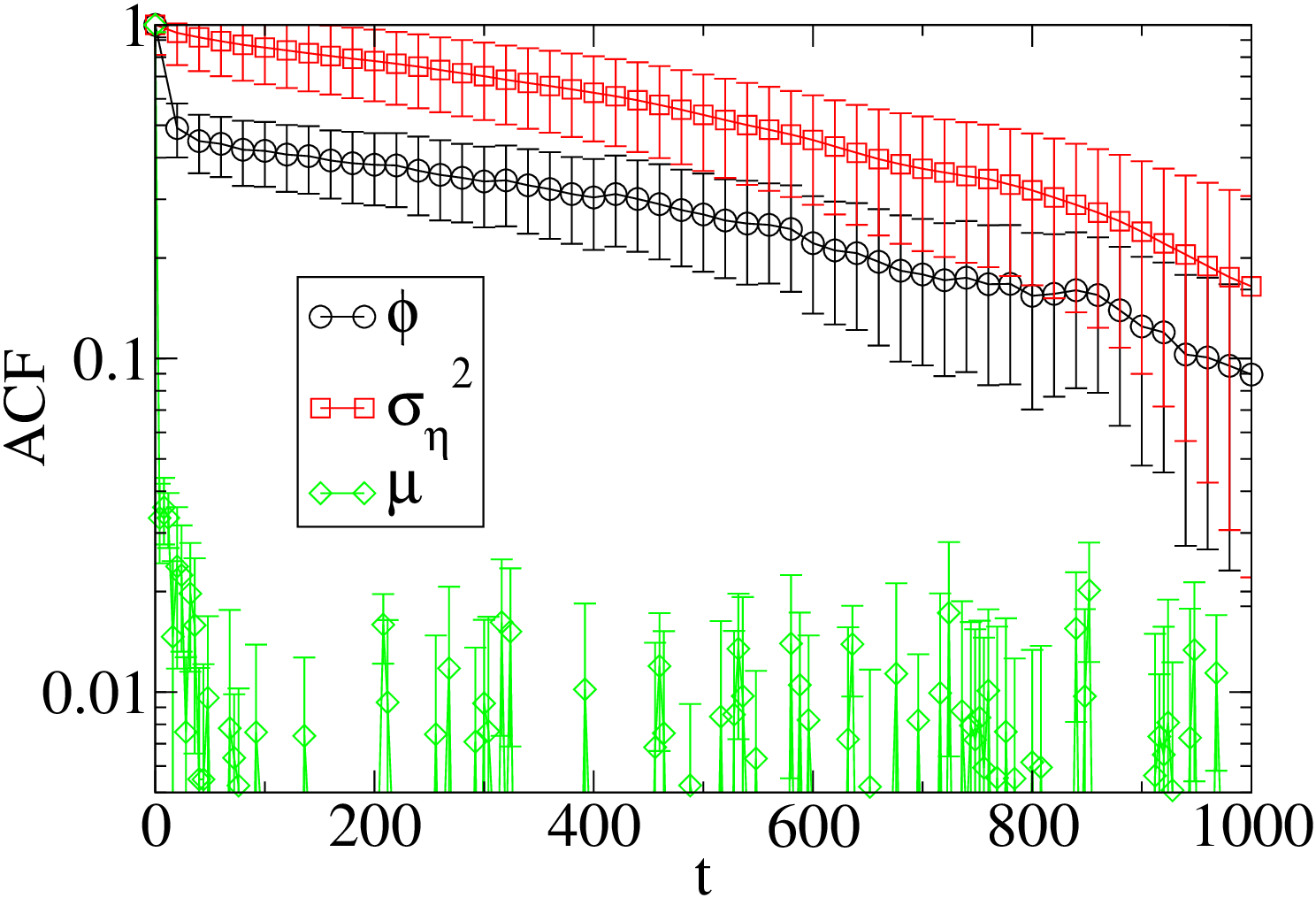}
\vspace{-2mm}
\caption{
Autocorrelation functions of the volatility variable $h_{100}$ (left)
and the sampled parameters (right). }
\label{fig:corr_nikkei}
\vspace{-2mm}
\end{figure}

In this section we make an empirical study of the SV model by the HMC algorithm.
The empirical study is based on daily data of the Nikkei 225 stock index.
The sampling period is  4 January 1995  to 30 December 2005 and
the number of the observations is 2706. 
Fig.\ref{fig:nikkei}(left) shows the time series of the data.
Let $p_i$ be the Nikkei 225 index at time $i$.
The Nikkei 225 index $p_i$ are transformed to returns as  
\be
r_i=100\ln(p_i/p_{i-1}-\bar{s}),
\label{eq:return}
\ee
where $\bar{s}$ is the average value of $\ln(p_i/p_{i-1})$. 
Fig.\ref{fig:nikkei}(right) shows the time series of returns calculated by
Eq.(\ref{eq:return}). 
We perform the same MCMC sampling by the HMC algorithm as in the previous section.
The first 10000 MC samples are discarded and then 20000 samples are recorded for the analysis. 
The ACF of sampled $h_{100}$ and sampled parameters are shown in Fig.\ref{fig:corr_nikkei}.
Qualitatively the results of the ACF are similar to those from the artificial financial data,
i.e. the ACF of the volatility and $\mu$ decrease quickly although the ACF of $\phi$ and
$\sigma_\eta^2$ decrease slowly.   
The estimated values of the parameters are summarized in Table 2.
The value of $\phi$ is estimated to be $\phi\approx 0.977$. This value is very close to one, which means 
the time series has the strong persistency of 
the volatility shock. 
The similar values are also seen in the previous studies\cite{SVMCMC1,SV}.

\begin{table}
\vspace{-1mm}
\centering
\begin{tabular}{c|cccc}
HMC         &  \makebox[20mm]{$\phi$}   & \makebox[20mm]{$\mu$}  &  \makebox[20mm]{$\sigma^2_{\eta}$}   &  \makebox[20mm]{$h_{100}$}    \\
  \hline
          &  0.977   & 0.52  &  0.020         &           \\
SD           &  0.006   & 0.13  &  0.005         &  \\
SE           &  0.001   & 0.0016&  0.001         &   \\
$2\tau_{int}$ &  560(190)    & 4(1)       &  1120(360)         &   21(5)             \\
\end{tabular}
\caption{Results estimated by the HMC for the Nikkei 225 index data.}
\vspace{-2mm}
\end{table}

\section{Conclusions}
We applied the HMC algorithm to the Bayesian inference of the SV model
and examined the property of the HMC algorithm in terms of the autocorrelation times of the sampled data. 
We observed that the autocorrelation times of the volatility variables and $\mu$ parameter are small.
On the other hand large autocorrelation times are observed for 
the sampled data of $\phi$ and $\sigma_\eta^2$ parameters.
The similar behavior for the autocorrelation times are also seen in the literature\cite{SV}.

From comparison of the HMC and Metropolis algorithms
we find that the HMC algorithm samples the volatility variables and $\mu$ more effectively 
than the Metropolis algorithm. 
However there is no significant difference for $\phi$ and $\sigma_\eta^2$ sampling.
Since the autocorrelation times of $\mu$ for both algorithms are estimated to be rather small 
the improvement of sampling $\mu$ by the HMC algorithm is limited.
Therefore the overall efficiency is considered to be similar to that of the Metropolis algorithm.

By using the artificial financial data we confirmed that 
the HMC algorithm correctly reproduces the true parameter values used to generate the artificial financial data.
Thus it is concluded that 
the HMC algorithm can be used as an alternative algorithm for the Bayesian inference of the SV model.

If we are only interested in parameter estimations of the SV model,
the HMC algorithm may not be a superior algorithm. 
However the HMC algorithm samples the volatility variables effectively.
Thus the HMC algorithm may serve as an efficient algorithm 
for calculating a certain quantity including the volatility variables. 

\subsubsection*{Acknowledgments.}
The numerical calculations were carried out on SX8 at the Yukawa Institute for Theoretical Physics in Kyoto University
and on Altix at the Institute of Statistical Mathematics.

{\it Note added in proof.} After this work was completed the author noticed a similar approach 
by Liu\cite{Liu}. 
The author is grateful to M.A.~Girolami for drawing his attention to this.


\begin{thebibliography}{99}
\bibitem{Stanley}
R.Mantegna and H.E.Stanley, 
{\it Introduction to Econophysics} (Cambride University Press, 1999).

\bibitem{CONT}
R.~Cont, Empirical Properties of Asset Returns: Stylized Facts and Statistical Issues,
{\it Quantitative Finance} {\bf 1} (2001) 223--236.

\bibitem{Stauffer}
D. Stauffer and T.J.P. Penna,
Crossover in the Cont-Bouchaud percolation model for market fluctuations,
{\it Physics A} {\bf 256} (1998) 284--290.

\bibitem{Lux}
T. Lux and M. Marchesi, Scaling and Criticality in a Stochastic Multi-Agent Model of a Financial Market
{\it Nature} {\bf 397} (1999) 498--500.

\bibitem{Iori}
G. Iori,
Avalanche Dynamics and Trading Friction Effects on Stock Market Returns,
{\it Int. J. Mod. Phys. C} {\bf 10} (1999) 1149--1162.

\bibitem{Stauffer2}
L.R. da Silva and D. Stauffer,
Ising-correlated clusters in the Cont-Bouchaud stock market model,
{\it Physics A} {\bf 294} (2001) 235--238.

\bibitem{Zhang}
D. Challet, A. Chessa, M. Marsili and Y-C. Zhang, From Minority Games to real markets,
{\it Quantitative Finance} {\bf 1} (2001) 168--176.
(2001)

\bibitem{AGENT}
M. Raberto, S. Cincotti, S.M. Focardi and M. Marchesi, 
Agent-based Simulation of a Financial Market,
{\it Physics A} {\bf 299} (2001) 319--327.

\bibitem{SPIN2}
S.~Bornholdt,  
Expectation Bubbles in a Spin Model of Markets: Intermittency from Frustration across Scales.
{\it Int. J. Mod. Phys. C} {\bf 12} (2001) 667--674.

\bibitem{SPIN4}
K.~Sznajd-Weron and R.~Weron, 
A Simple Model of Price Formation.
{\it Int. J. Mod. Phys. C} {\bf 13} (2002) 115--123.

\bibitem{Sanchez}
J.R. Sanchez, 
A Simple Model for  Stocks Markets,
{\it Int. J. Mod. Phys. C} {\bf 13} (2002) 639--644.

\bibitem{Yamano}
T. Yamano, 
Bornholdt's Spin Model of a Market Dynamics in High Dimensions,
{\it Int. J. Mod. Phys. C} {\bf 13} (2002) 89--96.

\bibitem{SPIN5}
T.~Kaizoji, S.~Bornholdt and Y.~Fujiwara, 
Dynamics of Price and Trading Volume in a Spin Model of Stock Markets with Heterogeneous Agents.
{\it Physica A} {\bf 316} (2002) 441--452.


\bibitem{SPIN6}
T.~Takaishi, 
Simulations of Financial Markets in a Potts-like Model,
{\it Int. J. Mod. Phys. C} {\bf 13} (2005) 1311--1317.

\bibitem{ARCH}
R.F.~Engle, 
Autoregressive Conditional Heteroskedasticity with Estimates of the Variance 
of the United Kingdom inflation,
{\it Econometrica} {\bf 60} (1982) 987--1007.

\bibitem{GARCH}
T.~Bollerslev, Generalized Autoregressive Conditional Heteroskedasticity,
{\it Journal of Econometrics} {\bf 31} (1986) 307--327.

\bibitem{EGARCH}
D.B.~Nelson, Conditional Heteroskedasticity in Asset Returns: A New Approach,
{\it Econometrica} {\bf 59} (1991) 347--370.

\bibitem{GJR}
L.R.~Glston, R.~Jaganathan and D.E.~Runkle,
On the Relation Between the Expected Value and the Volatility of the Nominal Excess on Stocks,
{\it Journal of Finance} {\bf 48} (1993) 1779--1801.

\bibitem{QGARCH1}
R.F.~Engle and  V.~Ng, Measuring and Testing the Impact of News on Volatility,
{\it Journal of Finance} {\bf 48} (1993) 1749--1778.

\bibitem{QGARCH2}
E.~Sentana, Quadratic ARCH models.
{\it Review of Economic Studies} {\bf 62} (1995) 639--661.


\bibitem{SVMCMC1}
E.~Jacquier, N.G.~Polson and P.E.~Rossi, Bayesian Analysis of Stochastic Volatility Models.
{\it Journal of Business \& Economic Statistics}, {\bf 12} (1994) 371--389.

\bibitem{SV}
S.~Kim, N.~Shephard and S.~Chib, Stochastic Volatility: Likelihood Inference and Comparison with ARCH Models,
{\it Review of Economic Studies} {\bf 65} (1998) 361--393.

\bibitem{SVMCMC2}
N.~Shephard and  M.K.~Pitt, 
Likelihood Analysis of Non-Gaussian Measurement Time Series,
{\it Biometrika} {\bf 84} (1997) 653--667.

\bibitem{Watanabe}
T.~Watanabe and Y.~Omori, A Multi-move Sampler for Estimating Non-Gaussian Time Series Models,
{\it Biometrika} {\bf 91} (2004) 246--248.

\bibitem{MCMCASAI}
M.~Asai, Comparison of MCMC Methods for Estimating Stochastic Volatility Models,
{\it Computational Economics} {\bf 25} (2005) 281--301.

\bibitem{HMC}
S.~Duane,  A.D.~Kennedy,  B.J.~Pendleton and D.~Roweth,
Hybrid Monte Carlo,
{\it Phys. Lett. B} {\bf 195} (1987) 216--222.

\bibitem{HMCGARCH}
T.~Takaishi, Bayesian estimation of GARCH model by Hybrid Monte Carlo,
{\it Proceedings of the 9th Joint Conference on Information Sciences 2006}, CIEF-214 \\
doi:10.2991/jcis.2006.159

\bibitem{ICIC2008}
T.~Takaishi, Financial Time Series Analysis of SV Model by Hybrid Monte Carlo,
{\it Lecture Notes in Computer Science} {\bf 5226} (2008) 929--936.

\bibitem{UKAWA}
A.~Ukawa, Lattice QCD Simulations Beyond the Quenched Approximation,
{\it Nucl. Phys. B (Proc. Suppl.)} {\bf 10} (1989) 66--145 

\bibitem{METRO}
N.~Metropolis {\it et al.}
Equations of State Calculations by Fast Computing Machines,
{\it J. of Chem. Phys.} {\bf 21} (1953) 1087--1091.

\bibitem{MH}
W.K~Hastings,  Monte Carlo Sampling Methods Using Markov Chains and Their Applications,
{\it Biometrika} {\bf 57} (1970) 97--109.

\bibitem{HOHMC}
T.~Takaishi, Choice of Integrators in the Hybrid Monte Carlo Algorithm,
{\it Comput. Phys. Commun.} {\bf 133} (2000) 6--17. 

\bibitem{HOHMC2}
T.~Takaishi, Higher Order Hybrid Monte Carlo at Finite Temperature,
{\it Phys. Lett. B} {\bf  540} (2002) 159--165.

\bibitem{MNHOMC}
T.~Takaishi and Ph.~de Forcrand,
Testing and Tuning Symplectic Integrators for Hybrid Monte Carlo Algorithm in Lattice QCD,
{\it Phys. Rev. E} {\bf 73} (2006) 036706.

\bibitem{Liu}
J.S.~Liu, {\it Monte Carlo Strategies in Scientific Computing}
(Springer, 2001).


\end{thebibliography}
\end{document}